\documentclass[12pt,oneside]{article}
\usepackage[left=3.4cm,top=3.6cm,bottom=3.6cm,right=3.4cm,head=0cm,foot=0.7cm]{geometry}
\usepackage{amsmath,amssymb}
\newcommand{\field}[1]{\mathbb{#1}}

\begin{document}
\title{Rethinking Boltzmannian Equilibrium}
\author{Professor Charlotte Werndl\\ (University of Salzburg and London School of Economics)\\and Professor Roman Frigg \\(London School of Economics)}
 \date{This article is forthcoming in:
Philosophy of Science http://journal.philsci.org/}

\maketitle
\begin{abstract}
Boltzmannian statistical mechanics partitions the phase space of a system into macro-regions, and the largest of these is identified with equilibrium. What justifies this identification? Common answers focus on Boltzmann's combinatorial argument, the Maxwell-Boltzmann
distribution, and maximum entropy considerations. We argue that they fail and present a new answer. We characterise equilibrium as the macrostate in which a system spends most of its time and prove a new theorem establishing that equilibrium thus defined corresponds to the largest macro-region. Our derivation is completely general, and does not rely on assumptions about the dynamics or internal interactions.
\end{abstract}

\newpage

\section{Introduction}

Boltzmannian statistical mechanics (BSM) partitions the phase space of a system into cells consisting of macroscopically indistinguishable microstates. These cells correspond to the macrostates, and the largest cell is singled out as the equilibrium macrostate. The connection is not conceptual: there is nothing in the \textit{concept} of equilibrium tying equilibrium to the largest cell. So what justifies the association of equilibrium with the largest cell?

After introducing BSM (Section 2), we discuss three justificatory strategies based on  Boltzmann's combinatorial argument, the Maxwell-Boltzmann distribution, and maximum entropy considerations, respectively. We argue that all three fail because they either suffer from internal difficulties or are restricted to systems with negligible interparticle forces. This prompts the search for an alternative answer. In analogy with the standard thermodynamic definition of equilibrium, we characterise equilibrium as the macrostate in which the system spends most of its time. We then present a new mathematical theorem proving that such an equilibrium macrostate indeed corresponds to the largest cell (Section 4). This result is completely general in that it is not based on any assumptions about the system's dynamics or the nature of interactions within the system.

\section{Boltzmannian Statistical Mechanics}\label{BSM}
Let us briefly introduce BSM.\footnote{For details see Frigg (2008, 103--21).} Consider a system consisting of $n$ particles which is isolated from the environment and in a bounded container. The system's state is specified by a point $x=(q,p)$ (the \textit{microstate}) in its $6n$-dimensional phase space $\Gamma$. The system's dynamics is determined by its classical Hamiltonian $H(x)$. Energy is preserved and therefore the motion is confined to the $6n-1$ dimensional energy hypersurface $\Gamma_{E}$ defined by $H(x)=E$, where $E$ is the energy value. The solutions of the equations of motion are given by the phase flow $\phi_{t}$ on $\Gamma_{E}$, where $\phi_{t}(x)$ is the state into which $x \in \Gamma_{E}$ evolves after $t$ time steps. $\Sigma_{E}$ is the Lebesgue-$\sigma$-algebra and, intuitively speaking, consists of all relevant subsets of $\Gamma_{E}$. $\Gamma$ is endowed with the Lebesgue measure $\mu$, which is preserved under $\phi_{t}$. This measure can be restricted to a measure $\mu_{E}$ on $\Gamma_{E}$ which is preserved as well and is normalised, i.e.\  $\mu_{E}(\Gamma_{E})=1$. $(\Gamma_{E},\Sigma_{E},\mu_{E},\phi_{t})$ is a \textit{measure-preserving dynamical system}.

Assume that the system can be characterised by a set $\{v_{1}, ..., v_{k}\}$ of \emph{macro-variables} ($k \in \field{N}$). The $ v_{i}$ assume values in $\field{V}_{i}$, and capital letters $V_{i}$ denote the values of $v_{i}$. A particular set of values $\{V_{1}, ..., V_{k}\}$ defines a \emph{macrostate} $M_{V_{1}, \ldots, V_{k}}$. We only write `$M$' rather than `$M_{V_{1}, ..., V_{k}}$' if the specific $V_{i}$ do not matter. A set of macrostates is complete iff (if and only if) it contains all states a system can be in.

A crucial posit of BSM is supervenience: a system's microstate uniquely determines its macrostate. Every macrostate $M$ is associated with a \textit{macro-region} $\Gamma_{M}$ consisting of all $x\in\Gamma_{E}$ for which the system is in $M$. For a complete set of macrostates the $\Gamma_{M}$ form a partition of $\Gamma_{E}$ (they do not overlap and jointly cover $\Gamma_{E}$).\footnote{For lack of space our focus is on the most common case where macrostates are defined relative to $\Gamma_{E}$. Our arguments generalise to cases in which the macrostates are defined relative to other subsets of $\Gamma$.}

The \emph{Boltzmann entropy} of a \emph{macrostate} $M$ is $S_{B}(M)\,:=\,k_{B}\log[\mu(\Gamma_{M})]$ ($k_{B}$ is the Boltzmann constant). The Boltzmann entropy of a \emph{system} at time $t$, $S_{B}(t)$, is the entropy of the system's macrostate at $t$: $S_{_{B}}(t):=S_{_{B}}(M_{x(t)})$, where $x(t)$ is the system's microstate at $t$ and $M_{x(t)}$ is the macrostate supervening on $x(t)$.

We denote the equilibrium macrostate by $M_{eq}$ and its macro-region by $\Gamma_{M_{eq}}$.
A crucial aspect of the standard presentation of BSM is that $\Gamma_{eq}$ takes up most of $\Gamma_{M_{E}}$. To facilitate the discussion, we introduce the term of $\beta$-dominance: $\Gamma_{M_{eq}}$ is \emph{$\beta$-dominant} iff $\mu(\Gamma_{M_{eq}}) \geq\beta$ for $\beta\in (1/2,1]$. Often equilibrium is characterised as a state where $\beta$ is close to one but nothing in what follows depends on a particular choice of $\beta$.

The characterisation of equilibrium as a $\beta$-dominant state goes back to Ehrenfest and Ehrenfest (1912, 30). While different versions of BSM explain the approach to equilibrium differently, $\beta$-dominance is a key factor in all of them. Those who favour an explanation based on ergodic theory have to assume that $\Gamma_{M_{eq}}$ takes up the majority of $\Gamma_{E}$ because otherwise the system would not spend most of the time in $\Gamma_{M_{eq}}$ (e.g.\ Frigg and Werndl 2011, 2012). Those who see the approach to equilibrium as the result of some sort of probabilistic dynamics assume that $\Gamma_{M_{eq}}$ takes up most of $\Gamma_{E}$ because they assign probabilities to macrostates that are proportional to $\mu(\Gamma_{M})$ and equilibrium comes out of as the most likely state only if the equilibrium macro-region is $\beta$-dominant (e.g.\ Boltzmann 1877). Proponents of the typicality approach see dominance as the key ingredient in explaining the approach to equilibrium and sometimes even seem to argue that systems approach equilibrium \textit{because} the equilibrium region takes up nearly all of phase space (e.g.\ Goldstein and Lebowitz 2004).
We do not aim to adjudicate between these different approaches. Our question is a more basic one: \emph{Why is the equilibrium state $\beta$-dominant}?

\section{Justificatory Strategies}\label{Justificatory Stratgies}

A look at the literature reveals three justificatory strategies. In practice these are often pursued side-by-side and seen as providing mutual support to each other. We will assess each of them and argue that none of them is conclusive.

\subsection{The Largest Number of Microstates and the Combinatorial Argument}\label{CA}

The leading idea of the first justificatory strategy is that \emph{equilibrium is the macrostate that is compatible with the largest number of microstates}. This strategy is exemplified by Boltzmann's (1877) combinatorial argument.\footnote{For details see Frigg (2008) and Uffink (2007).} The state of one particle is determined by a point in its $6$-dimensional state space $\Gamma_{\mu}$, and the state of a system of $n$ identical particles is determined by $n$ points in this space. Since the system is confined to a finite container and has constant energy $E$, only a finite part of $\Gamma_{\mu}$ is accessible. Boltzmann partitions the accessible part of $\Gamma_{\mu}$ into cells of equal size $\delta\omega$ whose dividing lines run parallel to the position and momentum axes. The result is a finite partition $\Omega := \{\omega_{1}, \, ..., \omega_{m}\}$, $m\in\field{N}$. The cell in which a particle's state lies is its \textit{coarse-grained microstate}. The coarse-grained microstate of the entire gas, called an \emph{arrangement}, is given by a specification of the coarse-grained microstate of each of particle.

The system's macro-properties depend only on how many particles there are in each cell and not on which particles these are. A specification of the `occupation number' of each cell is known as a \textit{distribution} $D=(n_{1}, n_{2},\ldots,n_{m})$ where $n_{i}$ is the number of particles whose state is in cell $\omega_{i}$. Since $m$ and $n$ are finite, there are only finitely many distributions  $D_1,\ldots,D_k$. Each distribution is compatible with several arrangements, and the number $G(D)$ of arrangements compatible with a given distribution $D=(n_{1}, n_{2},\ldots,n_{m})$ is
\begin{equation}\label{G(D)}
G(D)=\frac{n!}{n_1!n_2!\ldots,n_m!}.
\end{equation}

\noindent Every microstate $x$ of $\Gamma_{E}$ is associated with exactly one distribution $D(x)$. One then defines the set $\Gamma_{D}$ of all $x$ that are associated with a distribution $D$:
\begin{equation}\label{Gamma_D}
\Gamma_{D}= \{x\in\Gamma_E\,\,:\,\,D(x)=D\}.
\end{equation}

Since macro-properties are fixed by the distribution, distributions are associated with macrostates. So we ask: which of the distributions is the equilibrium distribution? Now Boltzmann's main idea enters the scene: equilibrium is the macrostate that is compatible with the largest number of microstates. To determine the equilibrium distribution, Boltzmann assumed that the energy $e_i$ of particle $i$ depends only on the cell in which it is located. Then the total energy is:
\begin{equation}\label{total}
\sum_{i=1}^{m}n_{i}e_{i}=E.
\end{equation}
He furthermore assumed that the number of cells in $\Omega$ is small compared to the number of particles (allowing him to use Stirling's formula). With the further trivial assumption that $\sum_{i=1}^{m}n_{i}=n$, Boltzmann shows that $\mu_{E}(\Gamma_{D})$ is maximal when
\begin{equation}\label{rig}
n_i = \gamma e^{\lambda e_{i}},
\end{equation}
where $\gamma$ and $\lambda$ are parameters which depend on $n$ and $E$. This is the \emph{discrete version of the Maxwell-Boltzmann distribution}. Thus the equilibrium macrostate corresponds to the Maxwell-Boltzmann distribution.\footnote{What $(\ref{rig})$ gives us is the distribution with the largest number of microstates (for the Lebesgue measure) on the $6N$-dimensional shell-like domain $\Gamma_{ES}$ specified by the condition (\ref{total}). It does \emph{not} give us the macro-region of maximal size (i.e., the distribution with the largest measure $\mu_{E}$ on the $6n-1$ dimensional $\Gamma_{E}$). As Ehrenfest and Ehrenfest (1959, 30) stress, the (not further justified) assumption is that the possible distributions and the proportion of the different distributions would not change if macrostates were instead defined on $\Gamma_{E}$.}

Its ingenuity notwithstanding, the combinatorial argument faces a number of important problems. The first is that it only applies to systems of non-interacting particles (Uffink 2007, 976-7). It provides a reasonable approximation for systems with negligible interparticle forces, but any other system is beyond its scope. SM ought to be a general theory of matter and so this is a serious limitation.

The second problem is the absence of a conceptual connection between equilibrium in thermodynamics (TD) and the idea that the equilibrium macrostate is the one that is compatible with the largest number of microstates. In TD equilibrium is defined as the state to which isolated systems converge when left to themselves and which they never leave once they have reached it. This has very little, if anything, in common with the kind of considerations underlying the combinatorial argument. This is a problem for anyone who sees BSM as a reductionist enterprise. While the precise contours of the reduction of TD to SM remain controversial, we are not aware of any contributors who maintain radical anti-reductionism.
Thus the disconnect between the two notions of equilibrium is a serious problem.

Two replies come to mind. The first points out that since $\Gamma_{M_{eq}}$ is the largest subset of  $\Gamma_{E}$, systems approach equilibrium and spend most of their time in $\Gamma_{M_{eq}}$. This shows that the BSM definition of equilibrium is a good approximation to the TD definition. This is not true in general. Whether a system spends most of its time in the $\beta$-dominant  $\Gamma_{M_{eq}}$ depends on the dynamics. If, for instance, the dynamics is the identity function, it is not true that a system out of equilibrium approaches equilibrium and spends most of its time there. The second reply points out that we know for independent reasons that the equilibrium distribution is the Maxwell-Boltzmann distribution. This argument will be discussed in the next subsection, and our conclusion will be guarded.

Finally, the combinatorial argument (even if successful) shows that the equilibrium macrostate is \textit{larger than any other macrostate}. However, as Lavis (2005) points out, this need not imply that the equilibrium is $\beta$-dominant. There may be a large number of smaller macrostates who \textit{jointly} take up a large part of $\Gamma_{E}$. So the combinatorial argument does in fact not show that equilibrium is $\beta$-dominant.

\subsection{The Maxwell-Boltzmann Distribution}\label{MB Distribution}
According to the next justificatory strategy, \emph{a system is in equilibrium when its particles approximately satisfy the Maxwell-Boltzmann distribution} (equation \ref{rig}))(e.g., Penrose 1989).
% $\chi_{_{V}}(x)$  is 1 if $q_{\mu}$ is in the container and $0$ otherwise.
This approach is misguided because the Maxwell-Boltzmann distribution is in fact the equilibrium distribution only for a limited class of systems, namely for systems consisting of particles with negligible interparticle forces. For particles with non-negligible interations different distributions correspond to equilibrium (Gupta 2002). Furthermore, for many simple models such as the Ising model (Baxter 1982) or the Kac-ring (Lavis 2008) the equilibrium macrostate also does not correspond to the Maxwell-Boltzmann distribution.

This is no surprise given that the two common derivations of the distribution in effect assume that particles are non-interacting. Boltzmann's (1877) derivation is based on equation (\ref{total}), the assumption that the total energy is the sum of the energy of the individual particles. This is true only if the particles are non-interacting, i.e. for ideal gases. While many expect that the argument also goes through for dilute gases (where this assumption holds approximately), the argument fails for non-negligible interactions. In Maxwell's 1860 derivation (see Uffink 2007) the non-interaction assumption enters via the postulate that the probability distributions in different spatial directions can be factorised, which is true only if there is no interaction between particles.

For these reasons the Maxwell-Boltzmann distribution is the equilibrium distribution only for a limited class of systems and cannot be taken as a general definition of equilibrium.

\subsection{Maximum Entropy}\label{Maximum Entropy}
A third stragey \emph{justifies dominance by maximum entropy considerations} along the following lines:\footnote{This strategy has been mentioned to us in conversation but it is hard to track down in print, at least in pure form. Albert's (2000) considerations concerning entropy seem to gesture in the direction of the third strategy.} we know from TD that, if left to itself, a system approaches equilibrium, and equilibrium is a maximum entropy state. Hence the Boltzmann entropy of a macrostate $S_{B}$ is maximal in equilibrium. Since $S_{B}$ is a monotonic function, the macrostate with the largest Boltzmann entropy is also the largest macrostate, which is the desired conclusion.

There are serious problems with the understanding of TD in this argument as well as with its implicit reductive claims. First, that a system, when left to itself, reaches equilibrium where entropy is maximal is often taken to be a consequence of the Second Law of TD, but it is not. As Brown and Uffink (2001) pointed out, that systems tend to approach equilibrium has to be added as an independent postulate, which they call the `Minus First Law'. But even if TD is amended with the Minus First Law, the conclusion does not follow. TD does not attribute an entropy to systems out of equilibrium. Thus, characterising the approach to equilibrium as a process of entropy increase \emph{is meaningless from a TD point of view}!

Even if all these issues could be resolved, there still would be a question why the fact that the TD entropy reaches a maximum in equilibrium would imply that the same holds for the Boltzmann entropy. To justify this inference, one would have to assume that the TD entropy reduces to the Boltzmann entropy. But it is far from clear that this is so. A connection between the TD entropy and the Boltzmann entropy has been established only for ideal gases, where the Sackur-Tatrode formula can be derived from BSM, which shows that both entropies have the same functional dependence on thermodynamic state variables. No such results are known for systems with iteractions. Furthermore, there are well-known differences between the TD and the Boltzmann entropy. Most importantly, the TD entropy is extensive while the Boltzmann entropy is not (Ainsworth 2012). But an extensive concept cannot reduce to a non-extensive concept (at least not without further qualifications).

For these reasons we conclude that maximum entropy considerations cannot be used to argue for the $\beta$-dominance of the equilibrium state.

\section{Rethinking Equilibrium}\label{Rethinking Equilibrium}
The failure of standard justificatory strategies prompts the search for an alternative answer. In this section we propose an alternative definition of equilibrium and introduce a new mathematical theorem proving that the equilibrium state thus defined is $\beta$-dominant.

The above strategies run into difficulties because there is no clear connection between the TD definition of equilibrium and  $\beta$-dominance. Our aim is to provide the missing connection by taking as a point of departure the standard TD definition of equilibrium and then exploiting supervenience to `translate' this macro-definition into micro-language.

The following is a typical TD textbook definition of equilibrium: `A thermodynamic system is in equilibrium when none of its thermodynamic properties are changing with time [...]' (Reiss 1996, 3) In more detail: equilibrium is the state to which an isolated system converges when left to its own and which it never leaves once it has been reached (Callender 2001; Uffink 2001). Equilibrium in TD is unique in the sense that the system always converges toward the \textit{same} equilibrium state. This leads to the following definition (the qualification `strict' will become clear later):

\begin{quote}
\textit{Definition 1: Strict BSM Equilibrium.} Consider an isolated system $S$ whose macrostates are specified in terms of the macro-variables $\{v_{1}, ..., v_{k}\}$, described as the measure-preserving dynamical system $(\Gamma_{E},\Sigma_{E},\mu_{E},\phi_{t})$. Let $M(x)$ be the macrostate that supervenes on microstate $x\in\Gamma$. Let $\Gamma_{M_{V_{1}, ..., V_{k}}}\,:=\,\{x \in \Gamma_E \, : \, M(x)=M_{V_{1}, ..., V_{k}} \}$ be the set of all microstates on which $M_{V_{1}, ..., V_{k}}$ supervenes. If there is a macrostate $M_{V_{1}^{*}, ..., V_{k}^{*}}$ satisfying the following condition, then it is the strict BSM equilibrium state of $S$: For \textit{all} initial states $x\in\Gamma_{E}$ at $t_{0}$ there exists a time $t^{*}$ such that $M_{V_{1}, ..., V_{k}}(\phi_{t}(x))=M_{V_{1}^{*}, ..., V_{k}^{*}}$ for all $t \geq t^{*}$, $i=1, ..., k$.\footnote{If one wants to avoid the $t^{*}$-dependence on the initial state, one can instead demand that there exists a time $t^*$ such that $v_{i}(t) = V_{i}^{*}$ for all initial states $M_{V_{1}, ..., V_{k}}$ and all $t \geq t^{*}$, $i=1, ..., k$.} We then write $M_{eq}\, := \, M_{V_{1}^{*}, ..., V_{k}^{*}}$.
\end{quote}
Note that this definition incorporates the Minus First Law of TD.

Before reflecting on this definition, we want to add a brief comment about reductionism. Reductive eliminativists may feel that a definition of equilibrium in SM that is based on `top down translation' of its namesake in TD undermines the prospect of a reduction of TD to SM. They would argue that equilibrium has to be defined in purely mechanical terms, and must then be shown to line up with the TD definition of equilibrium.

This point of view is not the only game in town and reduction can be had even if equilibrium is defined `top down' (as in the above definition).\footnote{For a discussion see Dizadji-Bahmani et al.\ (2010).} First, whether the above definition undercuts a reduction depends on one's concept of reduction. For someone with a broadly Nagelian perspective there is no problem: the above definition provides a bridge law, which allows the derivation of the requisite macro-regularities from the laws of the micro-theory.  And a similar argument can be made in the framework of New Wave Reductionism. Second, equilibrium is a macro-concept: when describing a system as being in equilibrium, we look at it in terms of macro-properties. From a micro-point of view there are only molecules bouncing around. They always bounce -- there is no such thing as a relaxation of particle motion to an immutable state. Hence a definition of equilibrium in macro-terms is no heresy.

Definition 1 is too rigid for two reasons. The first reason is Poincar\'{e} recurrence: as long as the `M' in SM refers to a mechanical theory that conserves phase volume (and there is widespread consensus that this is the case),\footnote{Hamiltonian Mechanics falls within this class, but the class is much wider.} any attempt to justify an approach to strict equilibrium in mechanical terms is doomed to failure. The system will at some point return arbitrarily close to its initial condition, violating strict equilibrium (Frigg 2008; Uffink 2007). The second reason is that such a justification is not only unattainable but also undesirable. Experimental results show that equilibrium is not the immutable state that classical TD presents us with because systems exhibit fluctuations away from equilibrium (Wang et al.\ 2002). Thus strict equilibrium is actually \emph{unphysical}.

Consequently, strict definitions of equilibrium are undesirable both for theoretical and experimental reasons. So let us relax the condition that a system has to remain in equilibrium for all $t \geq t^{*}$ by the weaker condition that it has has to be in equilibrium most of the time:
\begin{quote}
\textit{Definition 2: BSM $\alpha$-Equilibrium.}
Consider the same system as in Definition 1. Let $f_{M,x}(t)$ be the fraction of time of the interval $[t_{0}, \,t_{0}+t]$ in which the system's state is in $M$ when starting in initial state $x$ at $t_{0}$, and let $\alpha$ be a real number in $[0.5, 1]$. If there is a macrostate $M_{V_{1}^{*}, ..., V_{k}^{*}}$ satisfying the following condition, then it is the $\alpha$-equilibrium state of $S$: For \textit{all} initial states $x\in\Gamma_{E}$, $f_{M_{V_{1}^{*}, ..., V_{k}^{*}},x}(t) \geq \alpha$ in the limit $t \rightarrow \infty$. We then write $M_{\alpha\textnormal{-}eq}\, := \, M_{V_{1}^{*}, ..., V_{k}^{*}}$.
\end{quote}
An obvious question concerns the value of $\alpha$. Often the assumption seems to be that $\alpha$ is close to one. This is reasonable but not the only possible choice. For our purposes nothing hangs on a particular choice of $\alpha$ and so we leave it open what the best choice would be.

One last step is needed to arrive at the definition of equilibrium suitable for BSM. It has been pointed out variously that in SM, unlike in TD, we should not expect \textit{every} initial condition to approach equilibrium (see, for instance, Callender 2001). Indeed, it is reasonable to allow for a set of very small measure $\varepsilon$ for which the system does not approach equilibrium:
\begin{quote}
\textit{Definition 3: BSM $\alpha$-$\varepsilon$-Equilibrium.} Let $S$ and $f_{M,x}(t)$ be as above. Let $\alpha$ be a real number in $[0.5, 1]$; let $1>\varepsilon\geq 0$ be a small real number; and let $Y$ be a subset of $\Gamma_{E}$ such that $\mu_{E}(Y)\geq 1-\varepsilon$. If there is a macrostate $M_{V_{1}^{*}, ..., V_{k}^{*}}$ satisfying the following condition, then it is the $\alpha$-$\varepsilon$-equilibrium state of $S$: For \emph{all initial states $x\in Y$} $f_{M_{V_{1}^{*}, ..., V_{k}^{*}},x}(t) \geq \alpha$ in the limit $t \rightarrow \infty$. We then write $M_{\alpha\textnormal{-}\varepsilon\textnormal{-}eq}\, := \, M_{V_{1}^{*}, ..., V_{k}^{*}}$.
\end{quote}

Let us introduce the characteristic function of $\Gamma_{M}$, $1_{M}(x)$: $1_{M}(x)=1$ for $x\in \Gamma_{M}$ and $0$ otherwise. Definition 3 implies that for all $x \in Y$:\footnote{This shows that Defintion 4 is closely related to Lavis' (2005, 255) characterisation of TD-likeness.}
\begin{equation}\label{demanda}
\lim_{T\rightarrow\infty}\frac{1}{T}\int_{0}^{T}1_{M_{\alpha\textnormal{-}\varepsilon\textnormal{-}eq}}(\phi_{t}(x))dt\geq\alpha.
\end{equation}
An important assumption in this characterisation of equilibrium is that $\mu_{E}$ (and not some other measure) is the relevant measure. It is often argued that $\mu_{E}$ can be interpreted as a probability or typicality measure (Frigg and Hoefer 2013; Werndl 2013). The condition then says that the system's state spends more than a fraction $\alpha$ of its time in equilibrium with probability $1-\varepsilon$, or that typical initial conditions lie on trajectories which spend more than $\alpha$ of their time in equilibrium.

We contend that \emph{the relevant notion of equilibrium in BSM is $\alpha$-$\varepsilon$-equilibrium.} The central question then becomes: why is the $\alpha$-$\varepsilon$-equilibrium state $\beta$-dominant? Definition 3 in no way prejudges this question: it says nothing about the size of $\Gamma_{M_{\alpha\textnormal{-}\varepsilon\textnormal{-}eq}}$ nor does it in an obvious sense imply anything about it.

That $\Gamma_{M_{\alpha\textnormal{-}\varepsilon\textnormal{-}eq}}$ is $\beta$-prevalent follows from the following theorem, which we prove in the Appendix:
\begin{quote}
\emph{Equilibrium Theorem}: If $\Gamma_{M_{\alpha\textnormal{-}\varepsilon\textnormal{-}eq}}$ is an $\alpha$-$\varepsilon$-equilibrium of system $S$, then $\mu(\Gamma_{M_{\alpha\textnormal{-}\varepsilon\textnormal{-}eq}}) \geq \alpha(1-\varepsilon)$.
\end{quote}

We emphasise that the theorem is completely general in that no dynamical assumption is made (in particular it is not assumed that the system is ergodic). So the theorem also applies to strongly interacting systems such as solids and liquids.

The Equilibrium Theorem is the centre piece of our account. It shows in full generality that \textit{if the system $S$ has an $\alpha$-$\epsilon$-equilibrium, then the equilibrium state is $\beta$-dominant for $\beta\geq\alpha(1-\varepsilon)$.}\footnote{It is assumed that $\varepsilon$ is small enough so that $\alpha(1-\varepsilon)\geq 0.5$.} This provides the sought-after justification of the  $\beta$-dominance of the equilibrium state.

The Equilibrium Theorem makes the conditional claim that \textit{if} there is an $\alpha$-$\epsilon$-equilibrium, \textit{then} $\mu(\Gamma_{M_{\alpha\textnormal{-}\varepsilon\textnormal{-}eq}}) \geq \alpha(1-\varepsilon)$. As with all conditionals, the crucial and often vexing question is whether, and under what conditions, the antecedent holds. Some systems do not have equilibria. For instance, if the dynamics is given by the identity function, then no approach to equilibrium takes place, and the antecedent of the conditional is wrong. By contrast, epsilon-ergodicity allows for an equilibrium state to exist (Frigg and Werndl 2011). This raises the question under which circumstances the antecedent is true, which is an important question for future research.

\section{Conclusion}
Boltzmannian statistical mechanics partitions the phase space of a system into cells of macroscopically indistinguishable microstates. These cells are associated with the system's macrostates, and the largest cell is identified with equilibrium. What justifies the association of equilibrium with the largest cell? We discussed three justificatory strategies that can be found in the literature: that equilibrium is the macrostate compatible with the largest number of microstates, that equilibrium corresponds to the Maxwell-Boltzmann distribution and that most states are characterised by that distribution, and that equilibrium is the maximum entropy state. We argued that none of them is successful. This prompted the search for an alternative answer.  We characterised equilibrium as the state in which the system spends most of its time and presented a new mathematical theorem proving that such an equilibrium state indeed corresponds to the largest cell. This result is completely general in that it is not based on any assumptions either about the system's dynamics or the nature of interactions within the system. It therefore provides the first fully general justification of the claim that the equilibrium state takes up most of the accessible part of the system's phase space.

\pagebreak
\section*{Appendix: Proof of the Equilibrium Theorem}\label{A2}
The proof appeals to the ergodic decomposition theorem (cf.\ Petersen 1983, 81), stating that for a dynamical system $(\Gamma_{E},\Sigma_{E},\mu_{E},\phi_{t})$ the set $\Gamma_E$ is the disjoint union of sets $X_{\omega}$, each equipped with a $\sigma$-algebra $\Sigma_{X_{\omega}}$ and a probability measure $\mu_{\omega}$, and $\phi_{t}$ acts ergodically on each $(X_{\omega},\Sigma_{X_{\omega}},\mu_{\omega})$. The indexing set is also a probability space $(\Omega,\Sigma_{\Omega},P)$, and for any square integrable function $f$ it holds that:
\begin{equation}
\int_{\Gamma_{E}}fd\mu_{E}=\int_{\Omega}\int_{X_{\omega}}fd\mu_{\omega}dP.
\end{equation}
Application of the ergodic decomposition theorem for $f=1_{M_{\alpha\textnormal{-}\varepsilon\textnormal{-}eq}}(x)$ yields:
\begin{equation}\label{muede}
\mu_{E}(\Gamma_{M_{\alpha\textnormal{-}\varepsilon\textnormal{-}eq}})=\int_{\Gamma_{E}}1_{M_{\alpha\textnormal{-}\varepsilon\textnormal{-}eq}}(x)d\mu_{E}=
\int_{\Omega}\int_{X_{\omega}}1_{M_{\alpha\textnormal{-}\varepsilon\textnormal{-}eq}}(x)d\mu_{\omega}dP.
\end{equation}
For an ergodic system $(X_{\omega},\Sigma_{X_{\omega}},\mu_{\omega},\phi_{t})$ the long-run time average equals the phase average. Hence for almost all $x\in X_{\omega}$:
\begin{equation}\label{ergodic}
\lim_{T\rightarrow\infty}\frac{1}{T}\int_{0}^{T}1_{M_{\alpha\textnormal{-}\varepsilon\textnormal{-}eq}}(\phi_{t}(x))dt=
\int_{X_{\omega}}1_{M_{\alpha\textnormal{-}\varepsilon\textnormal{-}eq}}(x)d\mu_{\omega}=\mu_{\omega}(\Gamma_{M_{\alpha\textnormal{-}\varepsilon\textnormal{-}eq}}\cap X_{\omega}).
\end{equation}
From requirement (\ref{demanda}) and because $\phi_{t}$ acts ergodically on each $(X_{\omega},\Sigma_{X_{\omega}},\mu_{\omega})$, for almost all
$x\in X_{\omega}$, $X_{\omega}\subseteq Y$:
\begin{equation}
\alpha\leq\lim_{T\rightarrow\infty}\frac{1}{T}\int_{0}^{T}1_{M_{\alpha\textnormal{-}\varepsilon\textnormal{-}eq}}(\phi_{t}(x))dt=
\int_{X_{\omega}}1_{M_{\alpha\textnormal{-}\varepsilon\textnormal{-}eq}}(x)d\mu_{\omega}.
\end{equation}
Let us first consider the case $\varepsilon=0$, i.e. $\mu_{E}(Y)=1$. Here from equation~(\ref{muede}):
\begin{equation}
\mu_{E}(\Gamma_{M_{\alpha\textnormal{-}\varepsilon\textnormal{-}eq}})\geq\int_{\Omega}\alpha dP=\alpha.
\end{equation}
Hence if $\epsilon \geq 0$, it follows from equation~(\ref{muede}) that:
\begin{equation}
\mu_{E}(\Gamma_{M_{\alpha\textnormal{-}\varepsilon\textnormal{-}eq}})\geq\alpha(1-\varepsilon).
\end{equation}

\pagebreak
\section*{REFERENCES}

\noindent Ainsworth, Peter Mark 2012. ``Entropy in Statistical Mechanics.'' \emph{Philosophy of Science} 79: 542-560.\\

\noindent Albert, David 2000. \emph{Time and Chance}. Cambridge/MA and London: Harvard University Press.\\

\noindent Baxter, Rodney 1982. \emph{Exactly Solved Models in Statistical Mechanics}. San Diego: Academic Press Limited.\\

\noindent Boltzmann, Ludwig 1877. ``\"{U}ber die Beziehung zwischen dem zweiten Hauptsatze der mechanischen W\"{a}rmetheorie und der Wahrscheinlichkeitsrechnung resp.\ den S\"{a}tzen \"{u}ber das W\"{a}rmegleichgewicht.'' \emph{Wiener Berichte} 76: 373-435.\\

\noindent Brown, Harvey and Jos Uffink 2001. ``The Origins of Time-Asymmetry in Thermodynamics: The Minus First Law.'' \emph{Studies in History and Philosophy of Modern Physics} 32: 525-538.\\

\noindent Callender, Craig 2001. ``Taking Thermodynamics Too Seriously.'' \emph{Studies in History and Philosophy of Modern Physics} 32: 539-553.\\

\noindent Dizadji-Bahmani, Foad, Frigg, Roman and Stephan Hartmann 2010. ``Who Is Afraid of Nagelian Reduction?'' \emph{Erkenntnis} 73: 393-412.\\

\noindent Ehrenfest, Paul and Tatiana Ehrenfest 1959. \emph{The Conceptual Foundations of the Statistical Approach in Mechanics}. Ithaca, New York: Cornell University Press.\\

\noindent Frigg, Roman 2008. A Field Guide to Recent Work on the Foundations of Statistical Mechanics. In \emph{The Ashgate Companion to Contemporary Philosophy of Physics}, ed. Dean Rickles 99-196. London: Ashgate.\\

\noindent Frigg, Roman and Carl Hoefer 2010. Determinism and Chance from a Humean Perspective. In \emph{The Present Situation in the Philosophy of Science}, ed. Dieks, Dennis, Wencelao Gonzales, Stephan Hartmann, Marcel Weber, Friedrich Stadler und Thomas \"{U}bel, 351-372. Berlin and New York: Springer.\\

\noindent Frigg, Roman and Charlotte Werndl 2012. ``Demystifying Typicality.'' \emph{Philosophy of Science} 79:917-929.\\

\noindent Frigg, Roman and Charlotte Werndl 2011. ``Explaining Thermodynamic-Like Behaviour in Terms of Epsilon-Ergodicity.'' \emph{Philosophy of Science} 78: 628-652.\\

\noindent Goldstein, Sheldon and Joel L. Lebowtiz 2004. ``On the (Boltzmann) Entropy of Nonequilibrium
Systems.'' \emph{Physica D} 193:53-66.\\

\noindent Gupta, Mool C. 2003. \emph{Statistical Thermodynamics}. New Delhi: New Age International Publishing.\\

\noindent Lavis, David 2005. ``Boltzmann and Gibbs: An Attempted Reconciliation.'' \emph{Studies in
History and Philosophy of Modern Physics} 36: 245-273.\\

\noindent Lavis, David 2008. ``Boltzmann, Gibbs and the Concept of Equilibrium.'' \emph{Philosophy of Science} 75:682-696.\\

\noindent Penrose, Roger 1989. \emph{The Emperor's New Mind}. Oxford: Oxford University Press.\\

\noindent Petersen, Karl 1983. \emph{Ergodic theory}. Cambridge: Cambridge University Press.\\

\noindent Reiss, Howard 1996. \emph{Methods of Thermodynamics}. Mineaola/NY: Dover.\\

\noindent Uffink, Jos 2001. ``Bluff Your Way in the Second Law of Thermodynamics.'' \emph{Studies in History and Philosophy of Modern Physics} 32: 305-394.\\

\noindent Uffink, Jos 2007. Compendium of the Foundations of Classical Statistical
Physics. In \emph{Philosophy of Physics}, ed. Jeremy Butterﬁeld and John Earman,  923-1047. Amsterdam: North Holland.\\

\noindent Wang, Genmiao, Sevinck, Edith. M., Mittag, Emil, Searles, Debra J. and Denis J.\ Evans 2002. ``Experimental Demonstration of Violations of the Second Law of Thermodynamics for Small Systems and Short Time Scales.'' \emph{Physical Review Letters} 89:050601.\\

\noindent Werndl, Charlotte 2013. ``Justifying Typicality Measures in Boltzmannian Statistical Mechanics.'' \emph{Studies in History and Philosophy of Modern Physics} 44: 470-479.\\

\end{document}